\documentclass[10pt]{article}

\usepackage{graphicx}
\usepackage[english]{babel}
\usepackage{rotating}
\usepackage{amssymb}
\usepackage{amsmath, amssymb}
\usepackage{bbold}
\usepackage{graphicx}
\usepackage{dsfont}
\usepackage{natbib} % bibstyle chicago
\usepackage{url}
\usepackage{color}
\usepackage{colortbl}
\usepackage[font=footnotesize]{caption}
\usepackage{subcaption}
\usepackage{makecell}
\usepackage{booktabs, tabularx}
\usepackage{multirow}
\usepackage[autolanguage]{numprint}
\usepackage{bm}
\usepackage{changes} % \deleted
\usepackage{bbm}
\usepackage{authblk} % \author und affil
\usepackage{bibentry}
\usepackage{bbold}

\newcommand*\colvec[3][]{
    \begin{pmatrix}\ifx\relax#1\relax\else#1\\\fi#2\\#3\end{pmatrix}
}

\begin{document}
\title{A smooth dynamic network model for patent collaboration data}

\author[1]{Verena Bauer}%\thanks{Verena.Bauer@stat.uni-muenchen.de}}
\author[2]{Dietmar Harhoff}
\author[1]{G\"{o}ran Kauermann}
\affil[1]{Department of Statistics, Ludwig-Maximilians-Universit\"at M\"unchen, Germany.}

\affil[2]{Max Planck Institute for Innovation and Competition, Germany.}

\maketitle

\section*{Abstract}
The development and application of models, which take the evolution of network dynamics into account are receiving increasing attention. We contribute to this field and focus on a profile likelihood approach to model time-stamped event data for a large-scale dynamic network. We investigate the collaboration of inventors using EU patent data. As event we consider the submission of a joint patent and we explore the driving forces for collaboration between inventors. We propose a flexible semiparametric model, which includes external and internal covariates, where the latter are built from the network history.

\textit{{\bf{Keywords:}} profile likelihood, network data, event data, patent data, penalized spline smoothing, social network analysis}
\newpage

\section{Introduction}
The analysis of network data has seen increasing interest in the recent years. Many network data thereby contain a dynamic structure, be it the development of network ties over time or observations of the network at different time points. Such data structures have led to numerous extensions of classical network models.
A first paper in this direction is \cite{Robins:2001} who propose temporal dependence in an Exponential Random Graph Model (ERGM). The idea was generalized in \cite{Hanneke:2010} towards temporal Exponential Random Graph Models (tERGM). The principle idea behind the models is to include the network history as covariates in the model. This in turn forms a Markov Chain of networks. The model class has been extended and generalized in various ways. \cite{Leifeld:2018} focus on the implementation and added bootstrap methods for evaluating uncertainty. \cite{Krivitsky:2014} decomposed the network dynamics into the formation of new edges and the dissolution of existing edges leading to the separable temporal Exponential Random Graph Model (stERGM). 

A different strand of dynamic network models arise if time is considered as continuous. \cite{HollandLeinhardt:1977} develop a dynamic model for social networks based on a time-continuous Markov process. \cite{Snijders:2005} and \cite{Snijders:2010} extend this towards so-called stochastic actor-oriented models. The latter model is based on the assumption that the evolution of the network occurs as the consequence of small changes induced by the actors. It is further assumed that the observed network is derived from a Markov process evolving in continuous time, though the network is observed only at discrete time points. \cite{Greenan:2015} combines the approach with hazard function estimation and Cox regression models for duration time models \citep{Cox:1972}. A closely related model has been proposed by \cite{Butts:2008} for time-stamped relational data, defined as Relational Event Model (REM), which has been used in multiple applications, see e.g. \cite{Vu:2015, Vu:2017}. We also refer to \cite{StadtfeldBlock:2017} for extensions of this model class.
For time-stamped relational data estimation can be carried out using a partial likelihood approach. 
\cite{Perry:2013} estimate a Cox multiplicative intensity model for a directed e-mail network. \cite{Vu:2011} propose a continuous-time regression model for time-stamped network data. Estimation routines use an efficient partial likelihood approach focusing on large networks. 
This is also pursued in this paper. Instead of partial likelihood approaches one can also make use of complete likelihood estimation, see e.g. \cite{Stadtfeld:2011} or \cite{Butts:2017}.
A general discussion and comparison of different approaches in dynamic network modelling is found e.g. \cite{Block:2018} or \cite{Fritz:2020}.
Our approach is in line with the Relational Event Models, but we extend the model class by including non-linear time dynamics.
In this paper, we propose a profile likelihood approach for modeling time-stamped event data for large-scale network data. The data describe the collaboration of inventors based on joint patents. The successful submission of a new patent is thereby considered as the relational event and the number of joint patents of two inventors provides network based count data.

In the cited papers above, all covariate effects are included linearly in the model. We propose a semiparametric approach for modeling the covariates in a more flexible way. We follow the idea of penalized spline smoothing as proposed in \cite{Ruppert-etal:2003} \cite[see also][]{Eilers:1996, Ruppert-etal:2009}. The basic idea is to replace linear functions by spline based functions and to achieve smoothness, penalized spline smoothing can be considered as the state-of-the-art smoothing technique. We refer to \cite{Wood:2017} for a general discussion in the framework of (generalized) regression models.

The paper is organized as follows. In Section \ref{sec:patent_data} we introduce the patent data with some basic ideas and descriptive statistics. In Section \ref{sec:poisson_process}, we give an introduction to the notation and motivate the construction of the covariates from the network history.
We take a closer look on inference and derive how the model can be fitted based on a profile likelihood approach. This is extended to penalized spline smoothing. We give a brief outlook on computational issues, before we apply the proposed model in Section \ref{sec:data_analysis} to the example data. Finally, we summarize the most important issues.

\section{Patent data}
\label{sec:patent_data}
We will first introduce the patent data in detail before describing the model in the next section. We consider all patent applications submitted to the European Patent Office (EPO) and the German Patent and Trademark Office (Deutsches Patent- und Markenamt, DPMA), which listed at least one inventor with an address on German territory between 2000 and 2013. While this provides a comprehensive database of all inventions filed in patent applications by German inventors, we will restrict the subsequent analysis for the sake of space to two selected industrial areas, namely ``IT-methods'' as well as ``food chemistry''. Regarding the quality of the data we need to emphasize, that it is in principle possible that some inventors may have submitted applications directly to patent offices of other countries so that these are not in our database. In practice, however, such cases are extremely rare, since the invention would not enjoy patent protection in the inventors’ home country. The data were extracted from the PATSTAT database of the European Patent Office (version October 2018). For each patent we have information about the submission day (= time stamp) and for the majority of submission the inventors geographic coordinates of their registered home address at the time of submission is also given in the data. Apparently, the registered address might not be the work address, but still we consider it as allocation proxy which will be included as covariate subsequently. To do so we assume that the inventor location stays the same until new information due to new patent submissions is given. \\
The data structure is apparently of bipartite type, with inventors being connected through patents. In the subsequent analysis we focus on the relational aspect of the data by defining a relational event if two or more inventors submit a joint patent. This implies that single inventor submissions do not count as relational event while multi-inventor patent submissions lead to multiple relational events, all at the same time-point when submitting the patent. To make this point more clear, note that a patent with just two inventors corresponds to a single relational event (= one joint patent), while for instance a patent with three inventors leads to three pairwise relational ties (= three inventor pairs with a joint patent). The effect that multiple inventor patents will lead to multiple relational ties will be taken into account by an increased intensity for ties. Overall we take the inventors' point of view and consider all bilateral joint patents as events. We also excluded four patents which had more than 20 inventors. By doing so we also guarantee that our results are not overly influenced by a few patents with a large number of inventors. 

\begin{figure}[!ht]
	%\vspace{1cm}
	%\hspace{-2cm}
	\begin{minipage}{0.55\textwidth} 
		%\vspace{-2cm}
		\subcaptionbox{IT-methods: \\ Time period 2 (years 2005 - 2007)}
		{\vspace{-1cm}\includegraphics[width=1\textwidth]{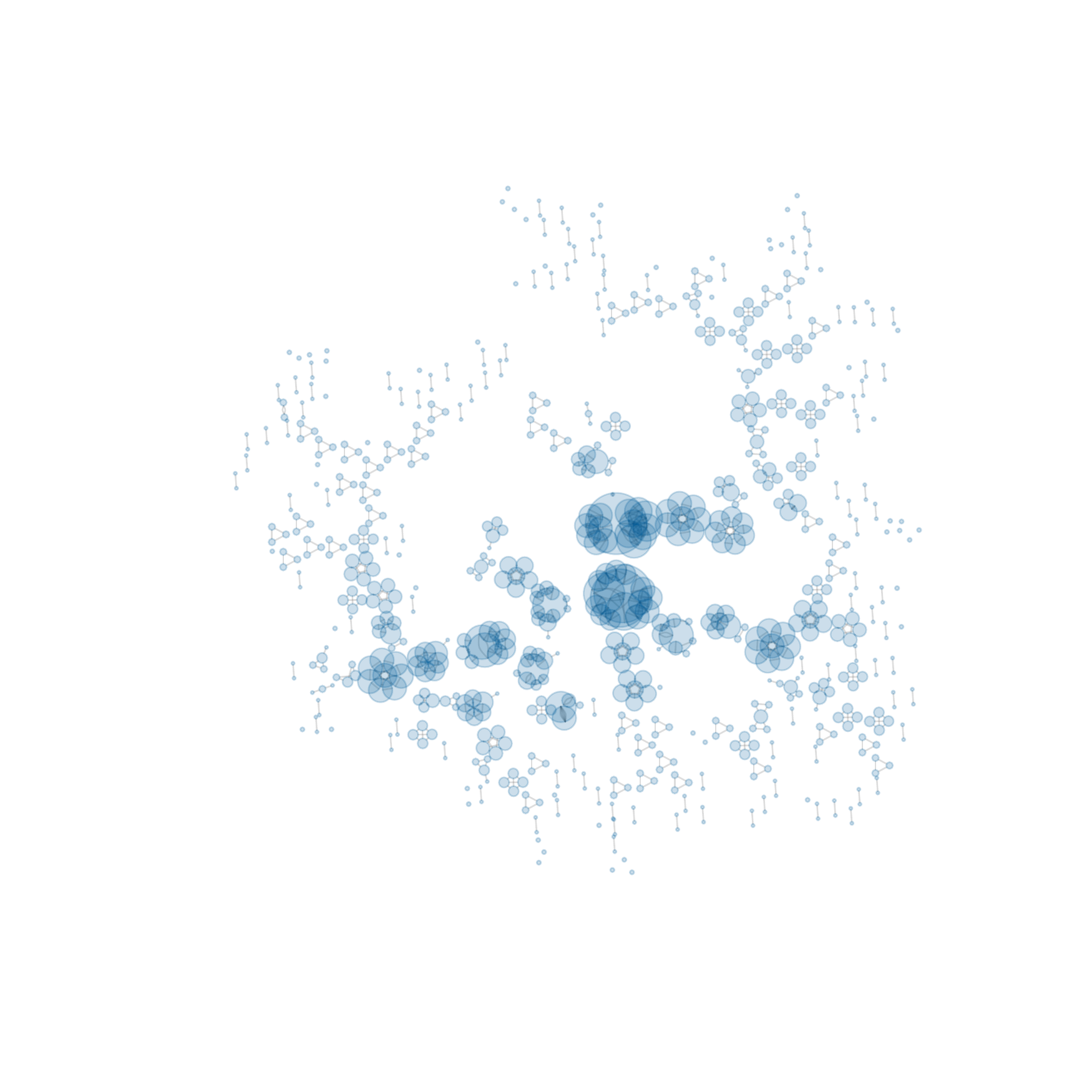}}
	\end{minipage}
	\hspace{-1cm}
	\begin{minipage}{0.55\textwidth} 
		%\vspace{-2cm}
		\subcaptionbox{IT-methods: \\ Time period 4 (years 2011 - 2013)}
		{\vspace{-1cm}\includegraphics[width=1\textwidth]{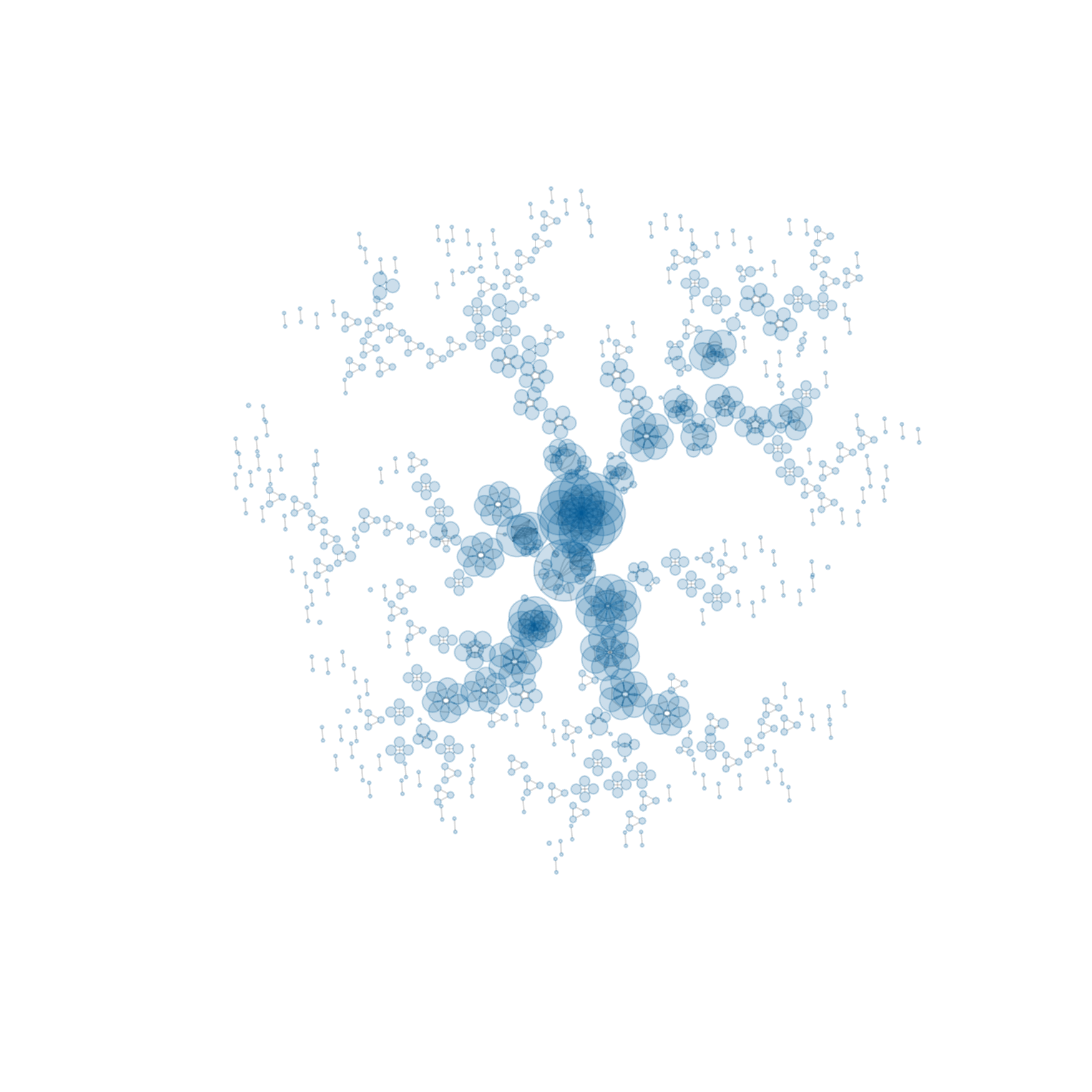}}
	\end{minipage}
	
	\begin{minipage}{0.55\textwidth} 
		%\vspace{-2cm}
		\subcaptionbox{food chemistry: \\ Time period 2 (years 2005 - 2007)}
		{\vspace{-1cm}\includegraphics[width=1\textwidth]{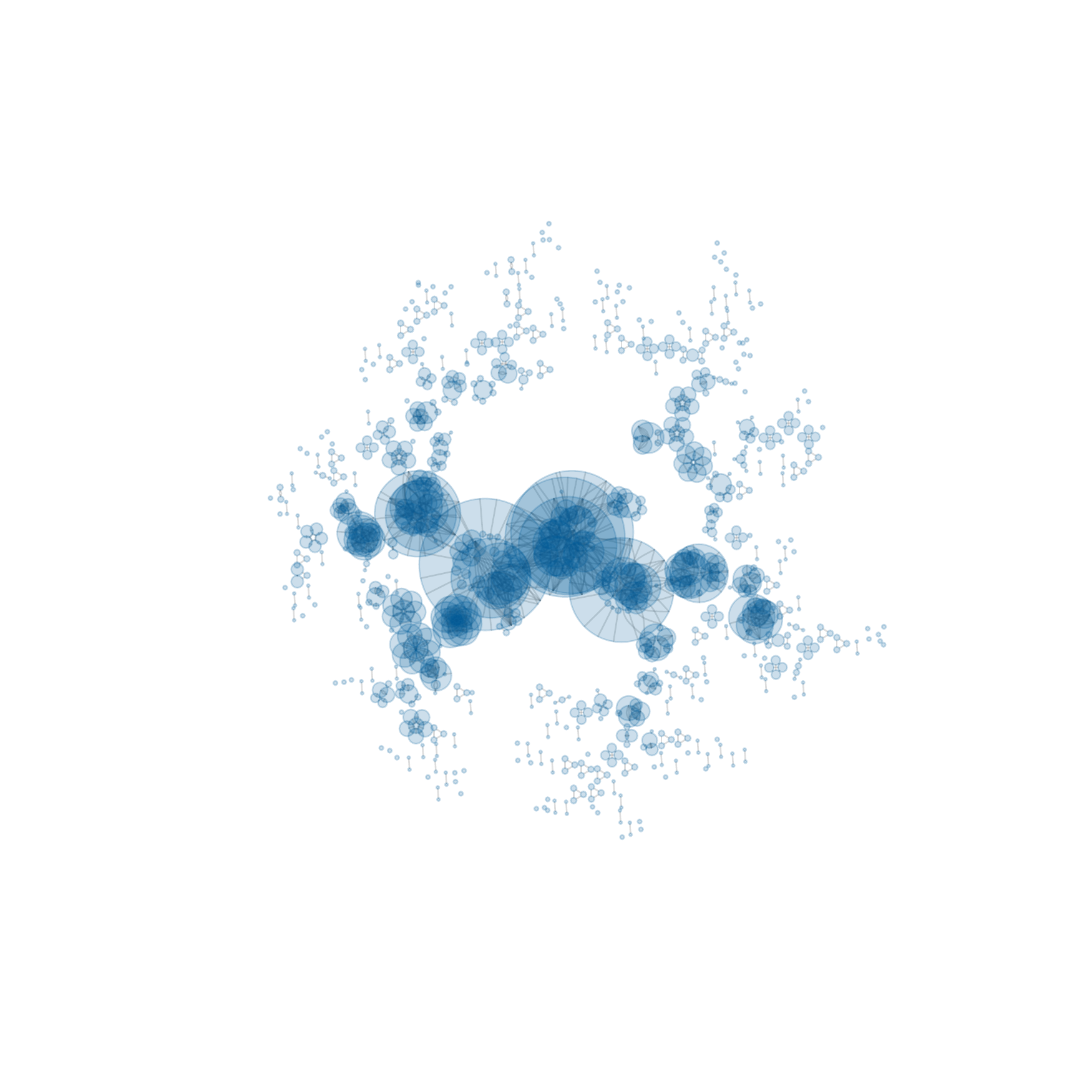}}
	\end{minipage}
	\hspace{-1cm}
	\begin{minipage}{0.55\textwidth} 
		%\vspace{-2cm}
		\subcaptionbox{food chemistry: \\ Time period 4 (years 2011 - 2013)}
		{\vspace{-1cm}\includegraphics[width=1\textwidth]{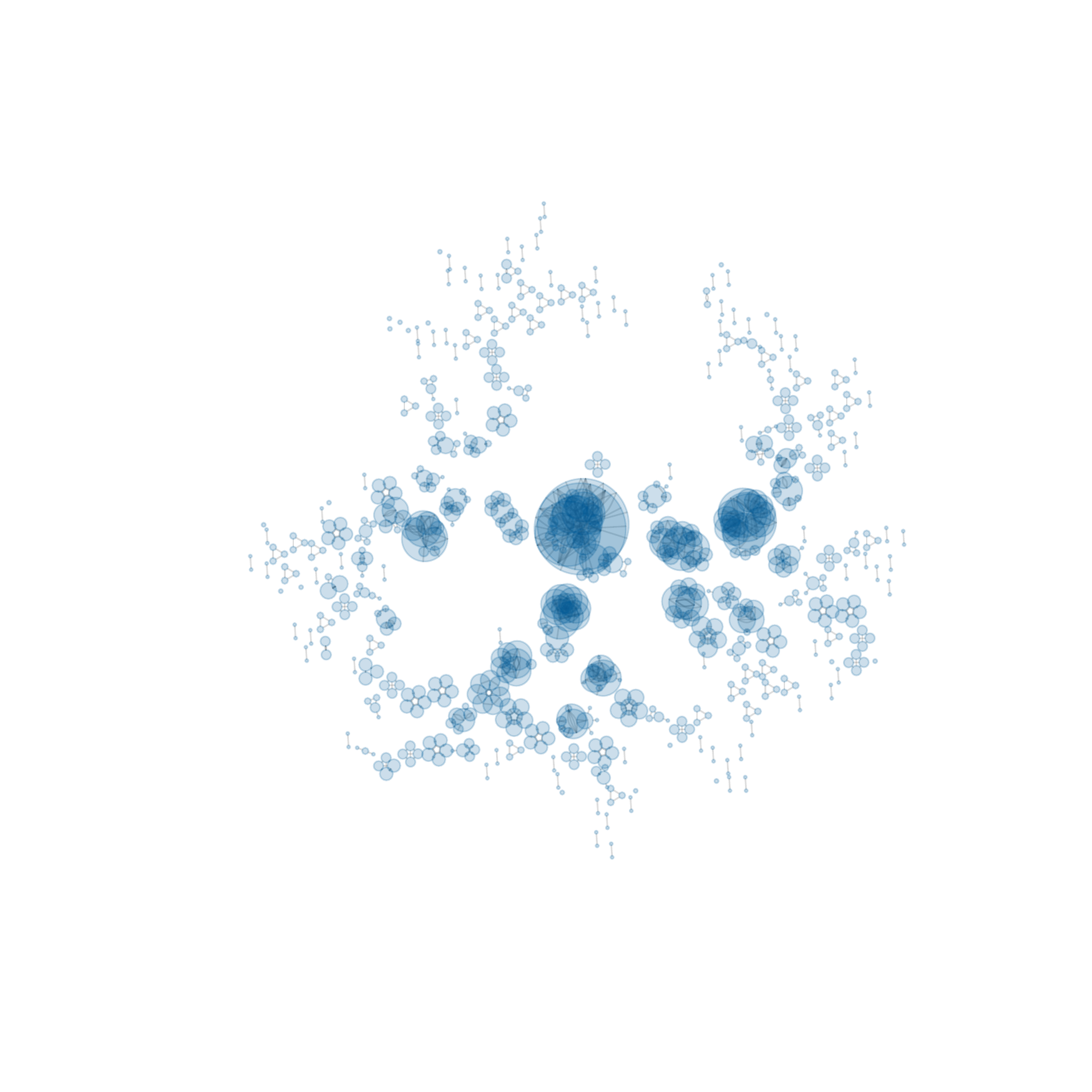}}
	\end{minipage}
	%\vspace{-1cm}
	\caption{Visualization of two time periods of the inventor network for  \textit{IT-methods} (107) and \textit{food chemistry} (118). Vertex size represent nodal degree. Colouring is transparent to better examine the clusters. The layout uses maximal connected components and applies the layout separately.} %Noticeable are that there are no loops due to the data set corresponds to the one used in the estimation where single ownership patents are ignored.}
	\label{Fig:network_graph_time_slot}
\end{figure}

We focus on two technological areas -- \textit{IT-methods} (classification number 107) and \textit{food chemistry} (classification number 118) -- with different numbers of inventors, patents and therefore network densities. 
% latex table generated in R 3.4.1 by xtable 1.8-3 package
% Mon Jun 17 09:23:57 2019
\begin{table}[t]
\centering
{\footnotesize
\begin{tabular}{llcc}
\toprule
& &IT-methods & food chemistry \\ 
\midrule
\midrule
  number of \ldots && &\\ 
\midrule
  \;\;\; \;\; inventors && 3480 & 2993 \\ 
  \midrule
  \;\;\; \;\; patents &&  1701 & 2078 \\ 
  \midrule
  \;\;\; \;\; single owner-ship patents & & 192  & 427 \\ 
  \midrule
  \;\;\; \;\; realized unique inventor pairs && 5525 & 5412 \\ 
  \midrule
    & min & 1 & 1 \\ 
  \;\;\; \;\; patents per inventor  & mean & 1.35 & 1.86 \\
  & max & 16 & 36 \\ 
  \midrule
      & min & 1 & 1 \\ 
  \;\;\; \;\; inventors per patent & mean & 2.76 & 2.68 \\  
  & max & 19 & 17 \\ 
\bottomrule
\end{tabular}
\caption{Summary statistics of two technological areas for the time period of 14 years. The statistics are summarized and averaged over time. } 
\label{tab:summary_stats}
}
\end{table}
Table \ref{tab:summary_stats} summarizes the selected inventor networks and Figure \ref{Fig:network_graph_time_slot} visualizes the network, separated for different time intervals. 
Compared to \textit{food chemistry} the \textit{IT-methods} technological area has a higher number of inventors, but a lower number of joint patents and single owner-ship patents. The number of patents per inventor is slightly higher for  food chemistry,  while the number of inventors per patent is about the same in the two fields. 

As time stamp we choose the earliest filing date, which is aggregated on a monthly basis. To adjust for incomplete data, we select only patents from the full years 2000 till the end of 2013, resulting in 168 months. We are interested in inventors that jointly apply for patents. Therefore, we only include inventors with at least one joint patent. Note, that there are of course single ownership patents in the data sets if the inventor also has joint patents. 

\begin{figure}[t]
\includegraphics[width=0.8\textwidth, page = 2]{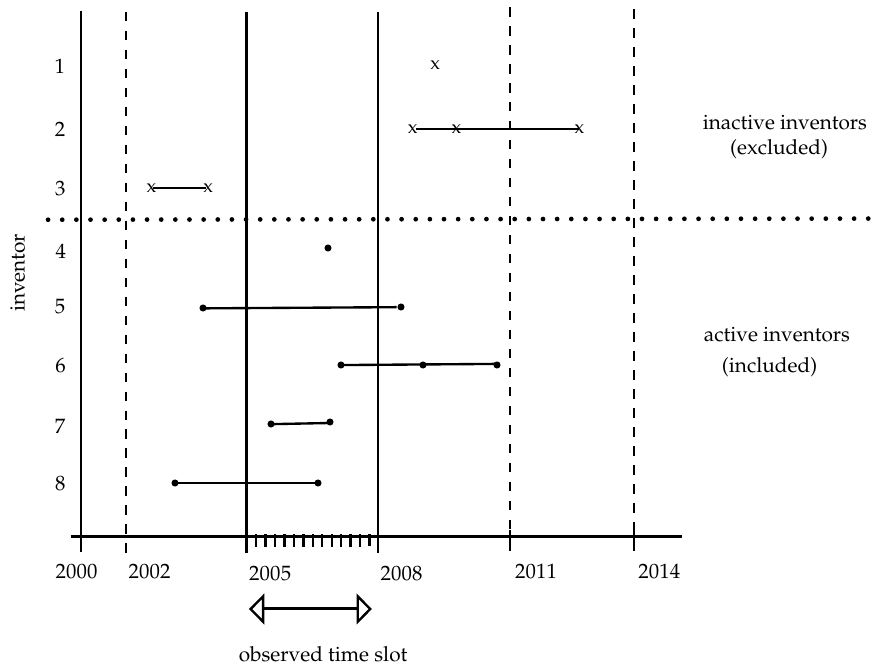}
\caption{Definition of active inventors. The time period from 2000 till the end of 2013 is divided in four periods ($2002-2004$, $2005-2007$, $2008-2010$ and $2011-2013$) of three years each. The data is aggregated on a monthly grid. The years 2000 and 2001 are used as a burn-in time.}
\label{Fig:time_slot}
\end{figure}

Noticeable is that the number of observed inventor pairs applying for a patent is quite small compared to the possible number of pairs $(N(N-1)/2)$. In other words the networks exhibit a low density, which is not uncommon in large networks. We aim to restrict the analysis to active inventors. To do so we divide the data into four periods, each of three years length.
We will analyse each time interval separately and include as inventors only those who are active within the considered period. We visualize our approach in Figure \ref{Fig:time_slot}. We include only active inventors in the option set. An active inventor is thereby defined as a person with at least one patent within the observed time period of three years (e.g. inventor 4 or 7 in Figure \ref{Fig:time_slot}), or at least one patent within and one beyond the time period (e.g. inventor 6 or 8 in Figure \ref{Fig:time_slot}), or at least one patent before and one after the time period (e.g. inventor 5 in Figure \ref{Fig:time_slot}).
The first two years of data from 2000 to the end of 2001 are used as "burn-in" period.
We also point out, that the covariates are based on a five years retrospective interval, meaning that the inventors' history beyond the five years is ignored in the calculation of the covariates. Table \ref{tab:summary_stats_time_slot} gives descriptive numbers of the network and the resulting covariates, which will be introduced later.

\begin{table}[t]
\centering
{\footnotesize
\begin{tabular}{llcc}
\toprule
%\multirow{3}{*}{\parbox{2cm}{}} & \multirow{3}{*}{\parbox{2cm}{}} & \multirow{3}{*}{\parbox{2cm}{Basic communication processes (105)}} & 
%\multirow{3}{*}{\parbox{2cm}{IT-Methods (107)}}  & \multirow{3}{*}{\parbox{2cm}{Analysis of biological material (111)}} & \multirow{3}{*}{\parbox{2cm}{ Food chemistry (118)}}\\
area && IT-methods  & food chemistry \\
\midrule
\midrule
no. of inventors  & & 767 - 900 & 753 - 949 \\ 
\midrule
no. of edges  &  & 993 - 1373 & 1188 - 1711 \\ 
\midrule
  density && 0.0033 - 0.004 & 0.0035 - 0.0042 \\ 
\midrule
\multirow{3}{*}{\parbox{2.5cm}{``\textit{patents\_ij}''}} 
& min & 0 - 0 & 0 - 0 \\ 
& mean & 1.38 - 1.77 & 2.11 - 2.64 \\ 
& max & 16 - 26 & 27 - 45 \\ 
 \midrule
\multirow{3}{*}{\parbox{2.5cm}{``\textit{joint\_patent}''}} 
& min & 0 - 0 & 0 - 0 \\
& mean & 0 - 0 & 0 - 0 \\ 
&max & 3 - 7 & 7 - 13 \\ 
\midrule
\multirow{3}{*}{\parbox{2.5cm}{``\textit{2-star}''}} 
& min & 0 - 0 & 0 - 0 \\ 
& mean & 3.06 - 3.85 & 4.11 - 4.93 \\ 
& max & 32 - 48 & 46 - 61 \\ 
\midrule
\multirow{3}{*}{\parbox{2.5cm}{``\textit{triangle}''}} 
& min & 0 - 0 & 0 - 0 \\ 
& mean & 0.01 - 0.01 & 0.01 - 0.01 \\ 
& max & 12 - 17 & 14 - 20 \\ 
\bottomrule
\end{tabular}
\caption{Summary statistics for the two technological areas.} 
\label{tab:summary_stats_time_slot}
}
\end{table}

%\clearpage

\section{Poisson process network model for count data}
\label{sec:poisson_process}
\subsection{Model description}
\label{sec:model_description}

We motivate the model by directly referring to our data example. Let $Z_r$ be a patent indexed with a running number $r = 1,...,R$. Each patent from one of the two considered technological areas can be defined through the following attributes:
\begin{itemize}
	\item $t_r$ = time point at which patent $r$ was successfully submitted
	\item $I_r$ = index list of inventors on patent $r$
	\item $z_r$ = additional covariates like geocoordinates of registered addresses of all inventors
\end{itemize}
For a set of actors (inventors) $A = \{1,...,N\}$ we define with $\boldsymbol{Y}(t) \in \mathbb{R}^{N \times N}$ the matrix valued Poisson process counting the number of (joint) patents. To be specific, let
\begin{align*}
	Y_{ij}(t) &= \text{cumulated number of joint patents of inventor $i$ and $j$}  \\
	&= \# \{r : (i,j) \in I_r, t_r \leq t, r = 1,...,R\}
\end{align*}
for $i,j = 1, ..., N$, where $Y_{ii}(t)$ defines the number of patents of inventor $i$ including single ownership patents. For each of the considered time intervals we set $t=0$  to mark the beginning of the three years period.  
For the network history we go back two years, that is we look at the process for $t \in [ -2, 3 ]$ measured in years, while the model is fitted to data for $t \in [0 , 3]$. We define with $Y_{ij,d} = Y_{ij} (t_{(d)})$ the evolving process, where $0\leq t_{(1)}, t_{(2)}, \ldots, t_{(m)} \leq 3 \mbox{ years}$ is the discretized version of time at which patents have been submitted. We model the intensity of the above process as
\begin{equation}
\lambda_0(t) \exp\left(x_{ij}(t) \beta\right)
\label{eq:intensity}
\end{equation}
where $\lambda_0(t)$ is the baseline intensity and $x_{ij}(t)$ is the covariate process, which will be defined in the following section.
We assume for simplicity that both, the baseline hazard as well as the covariate process are piecewise constant between the observed time points, that is
\begin{eqnarray*}
\lambda_0(t) =& \lambda_d &  \text{for} \;\; t \in ( t_{(d-1)}, t_{(d)}] \\
x_{ij}(t) =& x_{ij,d} & \text{for} \;\; t \in ( t_{(d-1)},  t_{(d)}].
\end{eqnarray*}
This leads to the log-likelihood function
\begin{align}
\label{eq:likeli_hazard}
l(\lambda_1, \ldots ,\lambda_m, \beta) = \hspace{9cm} \nonumber&\\
 \sum_{d=1}^{m} \left[ \sum_{(i,j) \in C_d} \left(\log \lambda_d + x_{ij, d} \beta \right)
 -\lambda_d\cdot \left( \sum_{(i',j') \in O_d} \exp \left( x_{i'j', d} \beta \right) \right) \right] &
\end{align}
where $C_d$ is the index set of events at time point $t_{(d)}$, $$ C_d = \{(i,j) : j > i; Y_{ij, d} > Y_{ij, d-1} \} $$  and $O_d$ is the ``option'' set, that is the set of inventor pairs that could submit a joint patent. This option set can be regarded as the set of inventors who are able to work together. In our application this restriction occurs from being in the same technological area and being an active inventor as defined above. Maximizing the above likelihood with respect to $\lambda_1, \ldots, \lambda_m$ yields
\begin{equation}
\hat{\lambda}_d = \frac{|C_d|}{\sum_{(i',j') \in O_d} \exp \left(x_{i'j', d} \beta\right)}
\label{eq:baselinefit}
\end{equation}
and inserting this in \eqref{eq:likeli_hazard} provides the profile log-likelihood
\begin{equation}
\label{eq:profile_likeli}
l(\beta) = \sum_{d=1}^{m} \left[ \sum_{(i,j) \in C_d}  x_{ij, d} \beta - |C_d| \log \left( \sum_{(i',j') \in O_d} \exp (x_{i'j', d} \beta) \right) \right],
\end{equation}
omitting all constant terms. Looking at (\ref{eq:baselinefit}) we want to point out that the baseline intensity takes into account that patents with multiple inventors lead to multiple relational events. As discussed above, a joint patent with two inventors gives one relational event, while a joint patent with three inventors already gives 3 relational events. Apparently, this is mirrored in the size $|C_d|$, meaning that the numerator in the baseline estimate in (\ref{eq:baselinefit}) adjusts for the multiplicity of relational events resulting 

In principle and based on the Poisson process we observe at each time point a single patent submission only, possibly with multiple authors. In our data, however, the time points are discretized so that at each discrete valued time point $t_{(d)}$ we may observe more than just one submitted patent.
Technically this is not a problem and does not require modifications, since in the case of multiple patent submissions the definition of the index set $C_d$ remains unchanged, but the index pairs in $C_d$ now refer to more than one patent submission. 
Again, the baseline estimate (\ref{eq:baselinefit}) is increased, this time due to multiple patents submitted at the same (discrete) timepoint. 
%Note that this in our data is majorly caused by one patent submitted at a discrete time point which has more than two inventors as patent holders. In this case, $|C_d = p \cdot (1-p)/2$ with $p$ being the number of inventors on the patent. As can be seen from the estimate of $\lambda_d$, the estimate of baseline intensity is increased, mirroring that with a multi-inventor patent the intensity of pairwise network links is increased as well. 
%or by more than one patent submitted at time point $t_{(d)}$ by different inventor pairs.

The above profile likelihood can also be motivated through  a partial likelihood approach, as shown subsequently. 
Let ${\bm{Y}}_d = (Y_{ij,d})$ be the process network matrix. We now assume that the probability for a single change $Y_{ij, d} = y_{ij, d-1} +1$ is proportional to 
$$ P(\bm{Y}_{d} = \bm{Y}_{d-1} + 1_{ij}) \propto \exp(x_{ij,d} \beta)$$
where $1_{ij}$ refers to an increment of $1$ in entry $Y_{ij,d}$ and  $x_{ij, d}$ is a vector of covariates calculated from the previous process matrix $\boldsymbol{Y}_{d-1}$. If $|C_d| = 1$, i.e. only a single patent with just two inventors was submitted by inventors $i$ and $j$ at time point $t_{(d)}$, we obtain
\begin{equation}
\label{eq:process_matrix_approx}
P(\bm{Y}_d|\bm{Y}_{d-1}) = \frac{\exp(x_{ij,d} \beta)}{\sum_{(i',j') \in O_d} \exp (x_{i'j', d} \beta)}.
\end{equation} 
If $|C_d| > 1$ we approximate \eqref{eq:process_matrix_approx} with

\begin{equation}
\label{eq:process_matrix_approx2}
	P(\bm{Y}_d | \boldsymbol{Y}_{d-1}) = \frac{\prod_{(i,j) \in C_d} \exp (x_{ij,d} \beta)}{\left[ \sum_{(i',j') \in O_d} \exp (x_{i'j',d} \beta)  \right] ^{|C_d|}}.
\end{equation}
Taking the logarithm we end up with the profile log likelihood given in \eqref{eq:profile_likeli}. We can now easily derive the log-likelihood from equation \eqref{eq:profile_likeli} and obtain the score function

\begin{equation*}
s(\beta) = \sum_{d=1}^{m} \left[\sum_{(i,j) \in C_d} x_{ij, d}^T - |C_d| \quad \frac{\sum_{(i',j') \in O_d} x_{i'j',d}^T \exp (x_{i'j', d} \beta)}{\sum_{(i',j') \in O_d} \exp (x_{i'j',d} \beta)}\right].
\end{equation*}
Defining
\[
\pi_{i'j',d}= \frac{\exp (x_{i'j',d} \beta)}{\sum_{(k',l') \in O_d}\exp (x_{k'l',d} \beta)}
\]
allows to write the second order derivative 
\begin{eqnarray*}
J(\beta) = - \sum_{d=1}^{m}|C_d| \left[ \sum_{(i',j') \in O_d} x_{i'j', d}^T x_{i'j',d} \pi_{i'j', d} \;- \hspace{4.5cm} \nonumber \right. \\  \left.  \left(\sum_{(i',j') \in O_d} x_{i'j', d}^T \pi_{i'j', d}\right)^T \left(\sum_{(i',j') \in O_d} x_{i'j', d}^T \pi_{i'j', d}\right) \right]. 
\end{eqnarray*}
In the survival model context, formula \eqref{eq:process_matrix_approx2} is also known as Breslow approximation \cite[see][]{Breslow:1974}.

\subsection{Covariates}
\label{sec:covariates}
\begin{figure}[t]
\begin{center}
\textsl{Toy network graph at time $t_{(d-1)}$}\par\medskip
\end{center}
\includegraphics[width=1\textwidth]{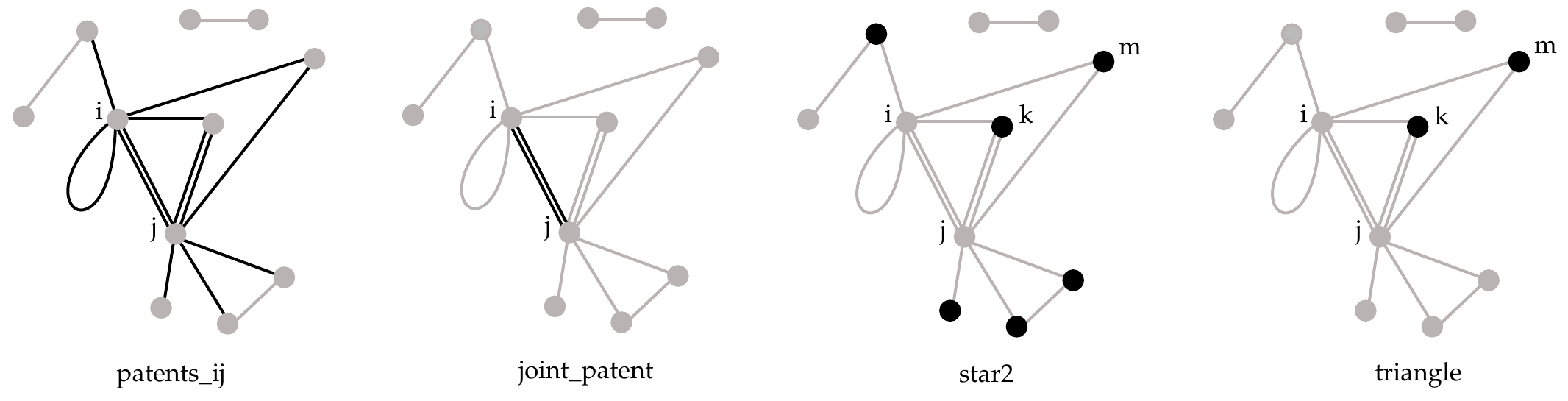}
\caption{Visualization of covariates from network history of a toy network graph:  Number of patents of inventor $i$ and $j$ with $x_{(1),ij,d} = 6 + 8$ (black edges), including self-loops (single ownership patents) and multiple patents (first panel). Number of joint patents of inventor $i$ and $j$ with $x_{(2),ij,d} = 2$ (black edges), counting the number of edges of $i$ and $j$ (second panel). Number of inventors that hold a joint patent with inventor $i$ or $j$ with $x_{(3),ij,d} = 3 + 5$ (black nodes in third panel). Number of inventors that jointly hold a patent with $i$ and $j$ with  $x_{(4),ij,d} = 2 $ (black nodes), counting $k$ twice because of a multi-patent (fourth panel).}
\label{Fig:statistics_patent_data}
\end{figure}

The covariate vector $x_{ij,d}$ is built from the network history itself as well as additional covariates. We define network specific covariates as endogenous, while the additional covariates are exogenous. We first describe network related covariates, which are described below and visualized in Figure \ref{Fig:statistics_patent_data}. Simple descriptive analyses are listed in Table \ref{tab:summary_stats_time_slot}. First, we take the total number of patents of inventor $i$ and $j$ at time point $t_{(d-1)}.$ That is 
\[x_{(1),ij,d} = Y_{ii,d-1} + Y_{jj,d-1}.\]
We refer to this quantity as ``\textit{patents\_ij}''.
Moreover, the number of previous ``\textit{joint\_patents}'' of inventor $i$ and $j$ is included as covariate, which is calculated through
\[x_{(2),ij,d} = Y_{ij,d-1}.\]
Furthermore, a so-called 2-star statistic (``\textit{2-star}'') is included, which expresses the number of inventors that hold a joint patent with inventor $i$ or $j$. This is obtained through
\[ x_{(3),ij,d} =
\sum_{\substack{k \neq i \\ k \neq j}} \mathbb{1}_{\{Y_{ik, d-1} > 0\}} + \sum_{\substack{k \neq j \\ k \neq i}} \mathbb{1}_{\{Y_{jk, d-1} > 0\}}.
\]
A common choice in network analysis are also ``\textit{triangle}'' statistics. This counts the number of inventors that jointly hold a patent with $i$ and $j$:
\[x_{(4),ij,d} =
\sum_{\substack{k \neq i \\ k \neq j}} \mathbb{1}_{\{Y_{ik, d-1} > 0\}} \cdot \mathbb{1}_{\{Y_{jk, d-1} > 0\}}.
\]
Note that the number of patents ($x_{(1)}$) as well as the number of joint patent holders ($x_{(2)}$) expresses the centrality of the inventors with respect to number of patents and number of collaborators, respectively. A summary of the distribution of the network related covariates is given in Table \ref{tab:summary_stats_time_slot}.

As exogenous covariates we include the inventor-pair-specific distance in kilometers, that is
\[
x_{(5),ij,d} = || s_{i,d} - s_{i,d}|| 
\] 
where $s_{i,d}$ are the geocoordinates of the address of inventor $i$ and $s_{j,d}$ accordingly and $|| \cdot ||$ denotes the Euclidean distance. We assume that the inventors do not move until new location information on the basis of submitting a new patent becomes available. %We expand the geographic coordinates of the first information given for the inventors to the past. 
To avoid leverage effects, we truncate distances over 1000 kilometers to 1000 kilometers.

\subsection{Semiparametric Estimation}
\label{sec:semiparametric_estimation}
We now extend the model towards penalized smoothing techniques to obtain more flexibility. We therefore replace the linear predictor $\eta_{ij,d} = x_{ij,d} \beta$ in \eqref{eq:profile_likeli} through the additive nonparametric setting
\begin{eqnarray*}
\eta_{ij,d} &=& m_{(1)}(x_{(1),ij,d}) + m_{(2)}(x_{(2),ij,d}) + \ldots. 
\end{eqnarray*}
Here $m_{(q)}(\cdot)$ are smooth but otherwise unspecified functions. To achieve identifiability of the model we postulate $m_{(q)}(0) = 0$ for $q>0$, which needs to be taken into account in the estimation. To estimate the unknown functions we employ B-splines and replace $m_{(q)}$ by 
\[
m_{(q)} = \sum_k B_{(q),k} u_{(q)}, %\sum_{q'=1}^q B^o_{q'}(x) u_{q'},
\]
where $B_{(q),k}$ is a $K$ dimensional B-spline basis spanning the observed range of covariate $x_{(q)}$. \cite[see][]{deBoor:1978,Wood:2017}. 

For simplicity of notation we now replace the index pair $(i,j)$ by a single index $l$ running from 1 to $n = \frac{N \cdot (N-1)}{2}$.
Consequently, we can rewrite
\begin{eqnarray*}
\eta_{l,d} &=& m_{(1)}(x_{(1),l,d}) + m_{(2)}(x_{(2),l,d}) + \ldots,
\end{eqnarray*}
which in matrix form leads to
\begin{eqnarray*}
\bm{\eta}_d &=& 
B_{(1),d} u_{(1)} + B_{(2),d} u_{(2)} + \ldots =
\bm{B}_d \bm{u}
\end{eqnarray*}
where $B_{(q),d}$ is the B-spline basis for the $q$-th covariate built from rows $B_{(q)}(x_{(q),l,d})$ for  $l = 1, \ldots, n$. Setting $\bm{B}_d = (B_{(1),d}, B_{(2),d}, \ldots)$ and $\bm{u}^T = (u^T_{(1)}, u^T_{(2)}, \ldots)$ provides the final notation.

With this notation we can reformulate the profile likelihood in \eqref{eq:profile_likeli} as:
\begin{eqnarray}
\sum_{d=1}^{m} \left[(\bm{B}_d\bm{u})^T \cdot \mathds{1}_{C_d} - |C_d| \cdot \log\left[\exp(\bm{B}_d\bm{u})^T \cdot \mathds{1}_{\lbrack n \times 1 \rbrack} \right] \right], 
\label{eq:likeli_smooth}
\end{eqnarray}
where $\mathds{1}_{C_d}$ is a vector defined as 
\[
\mathds{1}_{C_d} =
  \begin{cases}
   1,        & \text{if } l = (i,j) \in C_d \\
   0,        & \text{otherwise},
  \end{cases}
\]
$\mathds{1}_{\lbrack n \times 1 \rbrack}$ is a vector of ones of length $n$.

Following \cite{Eilers:1996} we use high dimensional bases but regularize the estimation by introducing a roughness penalty \cite[see also][]{Ruppert-etal:2003, Ruppert-etal:2009}. 
This leads to the penalized smooth log-likelihood
\begin{equation}
\label{eq:pen_smoooth_likeli}
l^{pen}(\bm{u},\lambda) = l(\bm{u}) - \frac{1}{2} \cdot \bm{u}^T \bm{K}(\bm{\lambda)} \bm{u},
\end{equation}
where $\bm{K}(\bm{\lambda})$ is a second-order penalty matrix. The smoothing parameter vector $\bm{\lambda}$ penalizes large differences in adjacent basis coefficients and can be estimated from the data. Details are provided in the Appendix C.

\subsection{Computational issues}
\label{sec:computational_issues}
In principle, computation is straight forward, because we can derive the corresponding likelihood function and its derivatives. One should bear in mind, though, we have a huge option set of pairs of inventors for each time point. A data set with $N$ inventors results in $N(N-1)/2$ times $T$ time points and therefore in about 18 million data points for e.g. $N = 1000$ inventors and $T=36$ months. This implies that estimation is numerically demanding, though feasible.

For estimating the parameters, we need to maximize the penalized smooth log-likelihood \eqref{eq:pen_smoooth_likeli} with its likelihood component defined in (\ref{eq:profile_likeli}). To do so, we can make use of the flexible toolbox available in the package \texttt{mgcv} \cite[see][for further information]{Wood:2011} in the software \texttt{R} \citep{RCore:2017}. This becomes possible by considering the data and the likelihood as "survival" data and applying proportional hazard models combined with a penalized Cox Model, which in turn results through a Poisson likelhood \cite[see][]{Whitehead:1980}. Estimation can therefore be carried out with standard routines after applying some data reorganization \cite[see][]{Tutz:2016}. At each event time $t_{(d)}$ an artificial response variable $y_{ij,d}$ for every inventor pair from the option set is included with $y_{ij,d}=1$ if a patent was submitted at time $t_{(d)}$ or $y_{ij,d} = 0$ if not.

\section{Data analysis}
\label{sec:data_analysis}
\begin{figure}[t]
\includegraphics[width=1\textwidth]{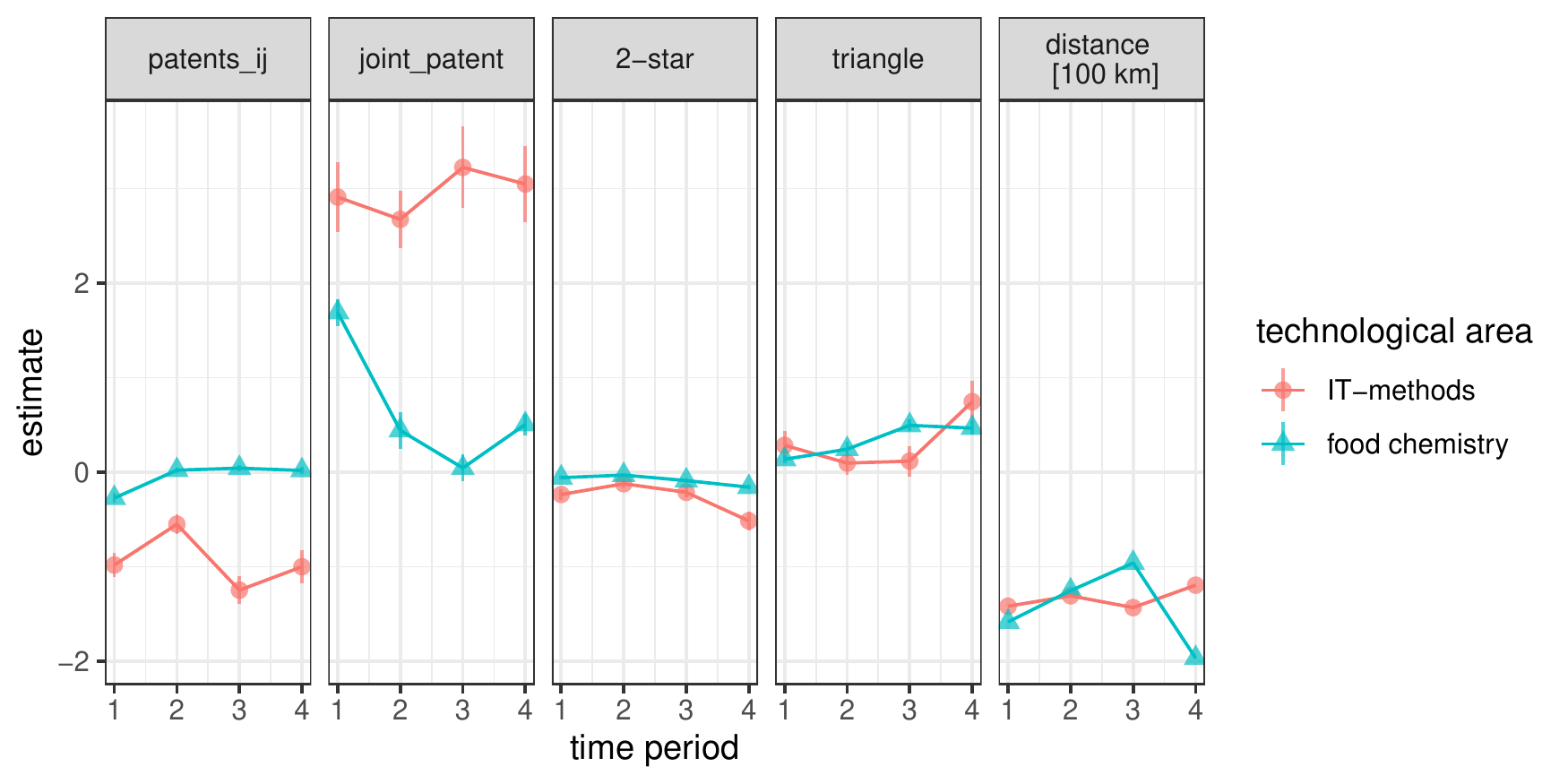}
\caption{Estimates for different covariates, technological areas and time periods. For each of the four areas and four covariates we have four estimates for the time periods with the corresponding errorbars (standard error $\times$ 2).}
\label{Fig:estimate_covariates_areas_period}
\end{figure}

We apply the proposed model to analyse the patent data described in Section \ref{sec:patent_data}. We start with a slightly simpler model than proposed and replace the smooth functions by simple linear functions. This easily allows to compare the effects for the two technology areas for the different time periods. All models include the above mentioned structural covariates ``\textit{patents\_ij}'', ``\textit{joint\_patent}'', ``\textit{2-star}'', and ``\textit{triangle}'', and the exogenous covariate ``\textit{distance [100 km]}''. 
Figure \ref{Fig:estimate_covariates_areas_period} compares the estimates for the four considered time periods. 
The different technology areas show more or less the same behaviour. The biggest difference can be seen for the variable \textit{joint\_patent}. The more joint patents two inventors have, the more likely they collaborate in the future. The estimates for 2-star and triangle are quite small. 
The distance in 100 kilometers has a negative effect on the patents meaning that inventors with regional proximity are collaborating more likely.

\begin{figure}[t]
\includegraphics[width=1\textwidth]{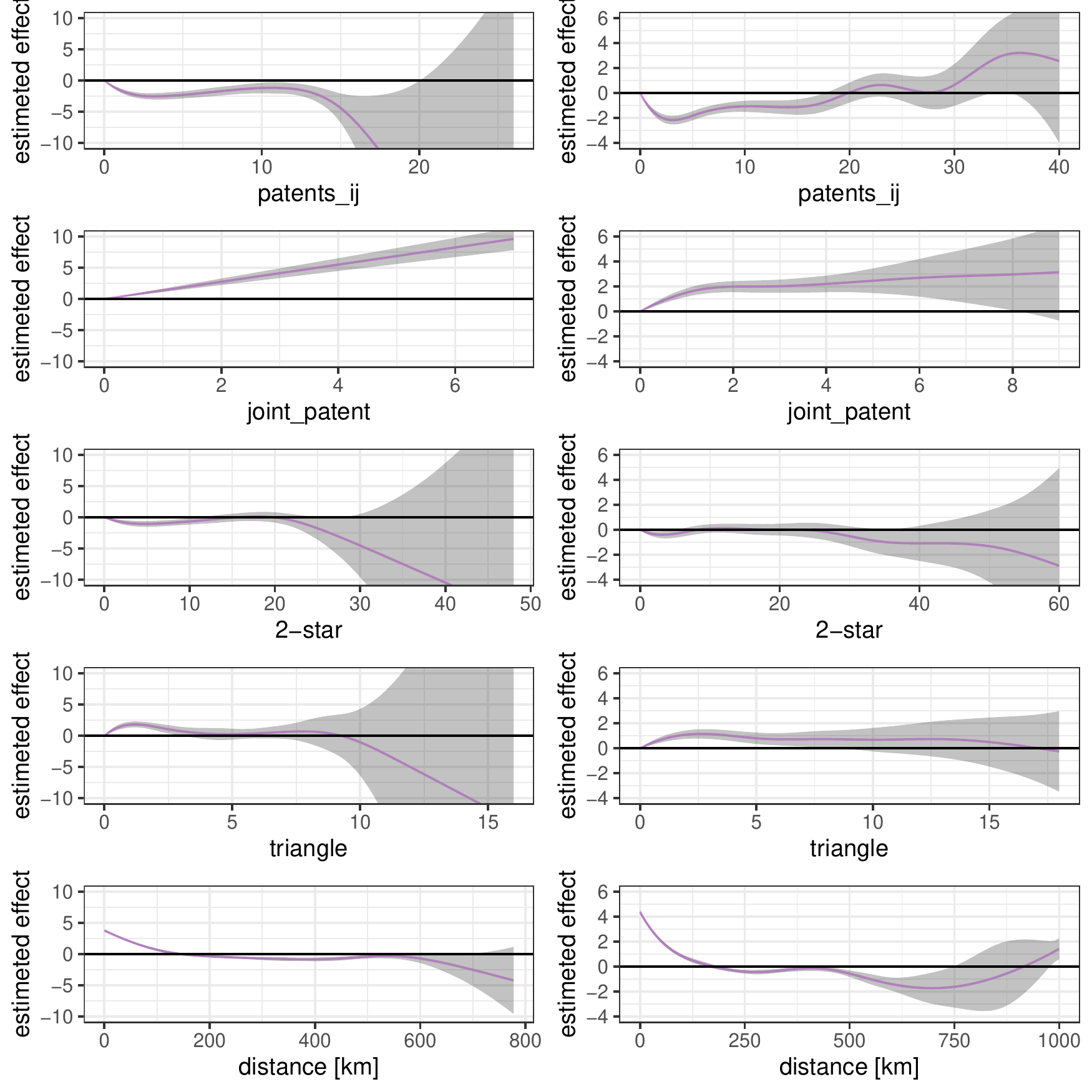}
\caption{Estimated smooth effects for \textit{IT-methods} (left panels) and \textit{food chemistry} (right panels) area and second time period.}
\label{Fig:all_smooth_effects_61_96_107_118}
\end{figure}

Next we explore the linearity and extend the model using smooth effects leading to semiparametric estimation with splines as proposed.
In Figure \ref{Fig:all_smooth_effects_61_96_107_118} we show exemplary for the second time period the fit of the model for the two technological areas. Estimates for the remaining time intervals can be found in the Appendix.
Form Figure \ref{Fig:all_smooth_effects_61_96_107_118} we see that the sum of patents of inventor $i$ and $j$ has a negative effect, whereas the number of joint patents has a positive and strong effect. This means that if the inventors have already submitted several own patents (with other inventors or even single inventor patents) their affinity of being involved in new patents decrease. On the other hand, if the inventor pair has already joint patents in the past, they are more likely to work together in future. The effect is nearly linear and stronger for the {\sl IT industry} compared to {\sl food and chemistry}.
The effect of the structural statistics like the number of inventors that hold a joint patent with inventor $i$ or $j$ (\textsl{2 star}), respectively, does not show a significant tendency. The effect of the number of inventors that jointly hold a patent with $i$ and $j$(\textsl{triangle}) has a small positive bounded influence, even though not that strong than the number of joint patents. Moreover, the geodesic distance of two inventors plays an important rule. There is a larger positive effect for small distances, which decreases with increasing distance. For distances larger than 250 kilometres the effect is almost zero or negative. This means that if there is a certain distance between the inventors, it does not matter how many kilometers exactly.

\begin{figure}[t]
\includegraphics[width=1\textwidth]{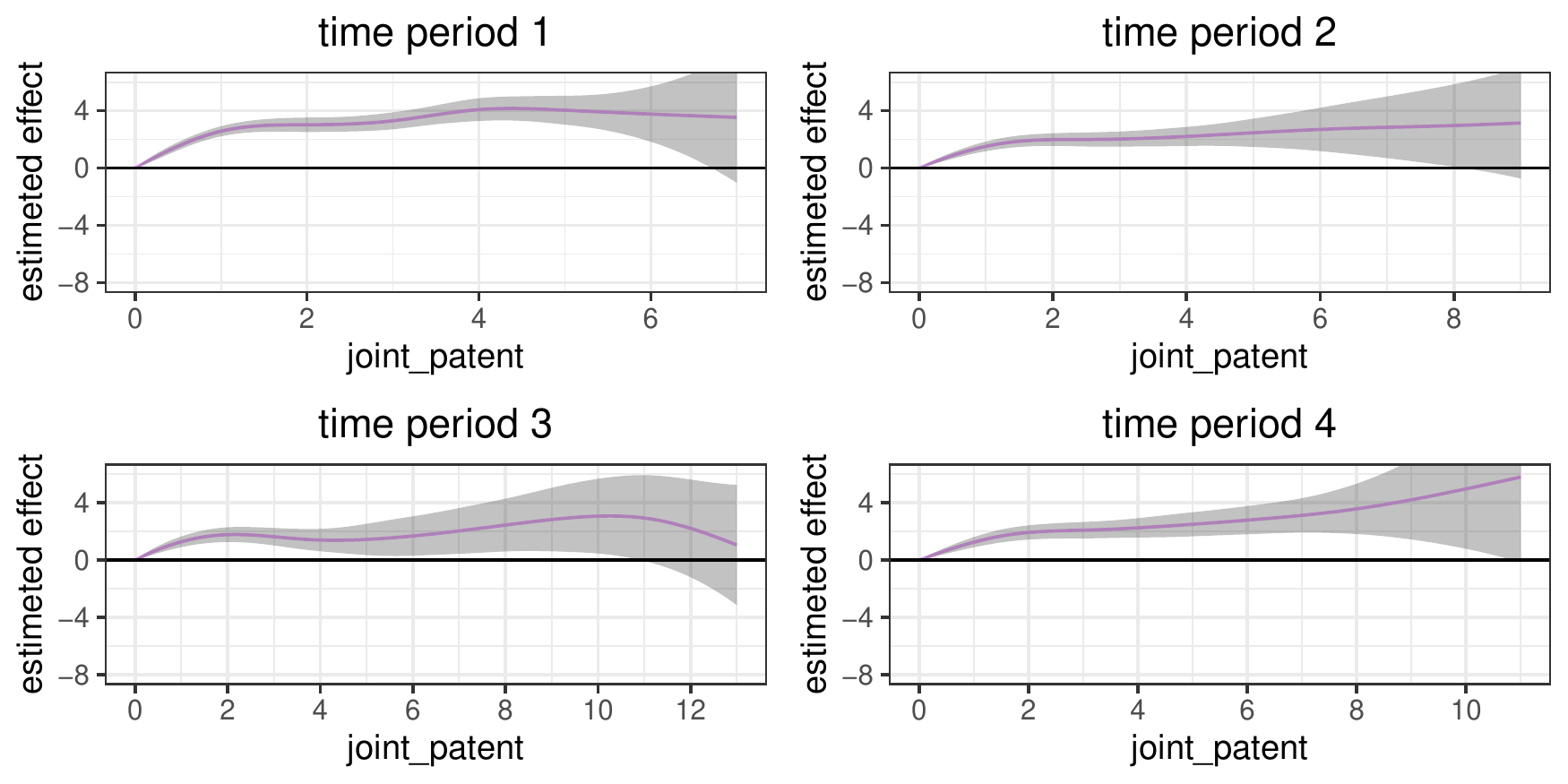}
\caption{Estimated smooth effects for \textit{``joint\_patent''} of \textit{food chemistry} (118) area and different time periods.}
\label{Fig:smooth_effect_joint_patent_118}
\end{figure}

Figure \ref{Fig:smooth_effect_joint_patent_118} visualizes the positive effects of \textit{``joint\_patent''} for the four time periods exemplary for the {\sl food chemistry} area.
Each time period lasts 36 months. The tendency of the effects is about the same for all periods; there is a steep increase at the beginning, which then becomes bounded. In period three and four the effect decreases and increases, respectively, at the end of the observation period. This should not be interpreted too strictly as the frequency of more than 10 joint patents is quite low.
We can see similar behaviours for the other areas (see Appendix).

\section{Conclusion}
\label{sec:discussion}

In this paper we propose a flexible approach to model large-scale dynamic network data with structural and exogenous covariates. Our approach is based on a profile likelihood method exploiting well-established estimation routines. We apply this idea to a large data set of patents submitted jointly by inventors from Germany between 2000 and 2013. We show advantages of including covariates in a semiparametric and therefore flexible way. The results show the driving forces in collaboration of inventors and demonstrate their behaviour over time. The models can be fitted with standard software employing the link to the Cox model and therefore invite to be used in other data constellations as well.

\bibliographystyle{spbasic} 
\bibliography{literature}  

\newpage

\begin{appendix}
\section*{Appendix A: Further Results}
\label{sec:appendixResults}

\begin{figure}[!ht]
\includegraphics[width=1\textwidth]{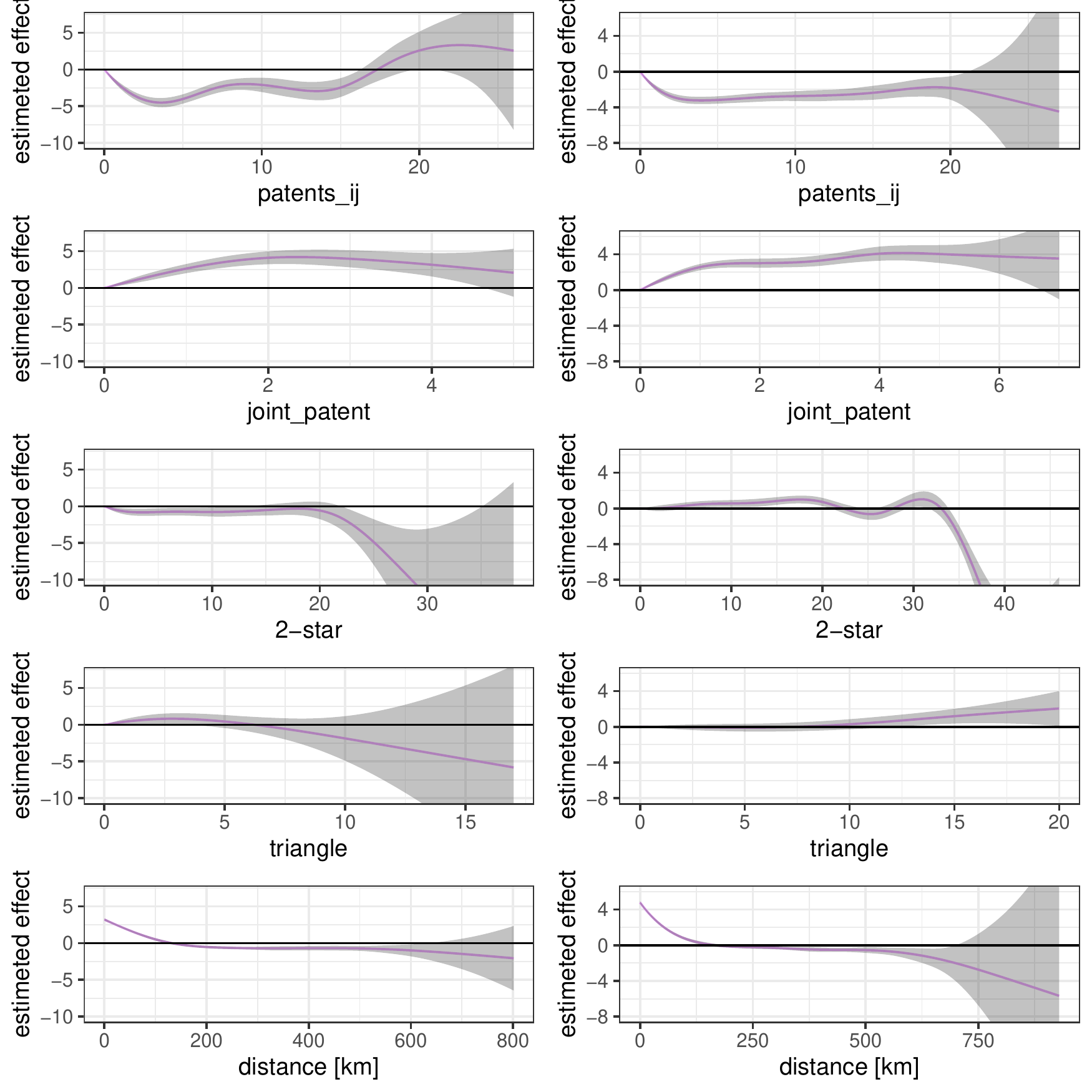}
\caption{Estimated smooth effects for \textit{IT-methods} (left panels) and \textit{food chemistry} (right panels) area and first time period.}
\label{Fig:all_smooth_effects_25_60_107_118}
\end{figure}

\begin{figure}[!ht]
\includegraphics[width=1\textwidth]{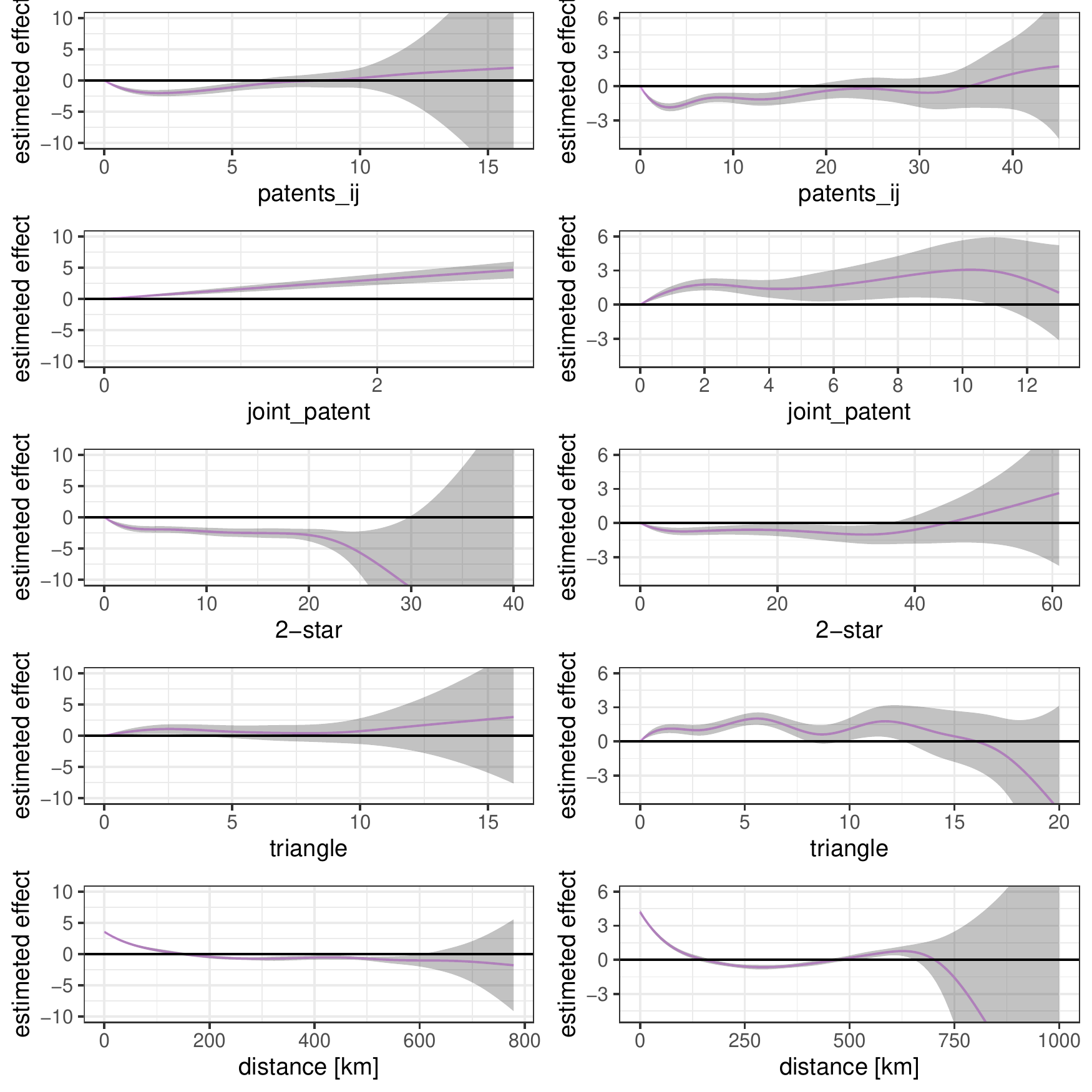}
\caption{Estimated smooth effects for \textit{IT-methods} (left panels) and \textit{food chemistry} (right panels) area and third time period.}
\label{Fig:all_smooth_effects_97_132_107_118}
\end{figure}

\begin{figure}[!ht]
\includegraphics[width=1\textwidth]{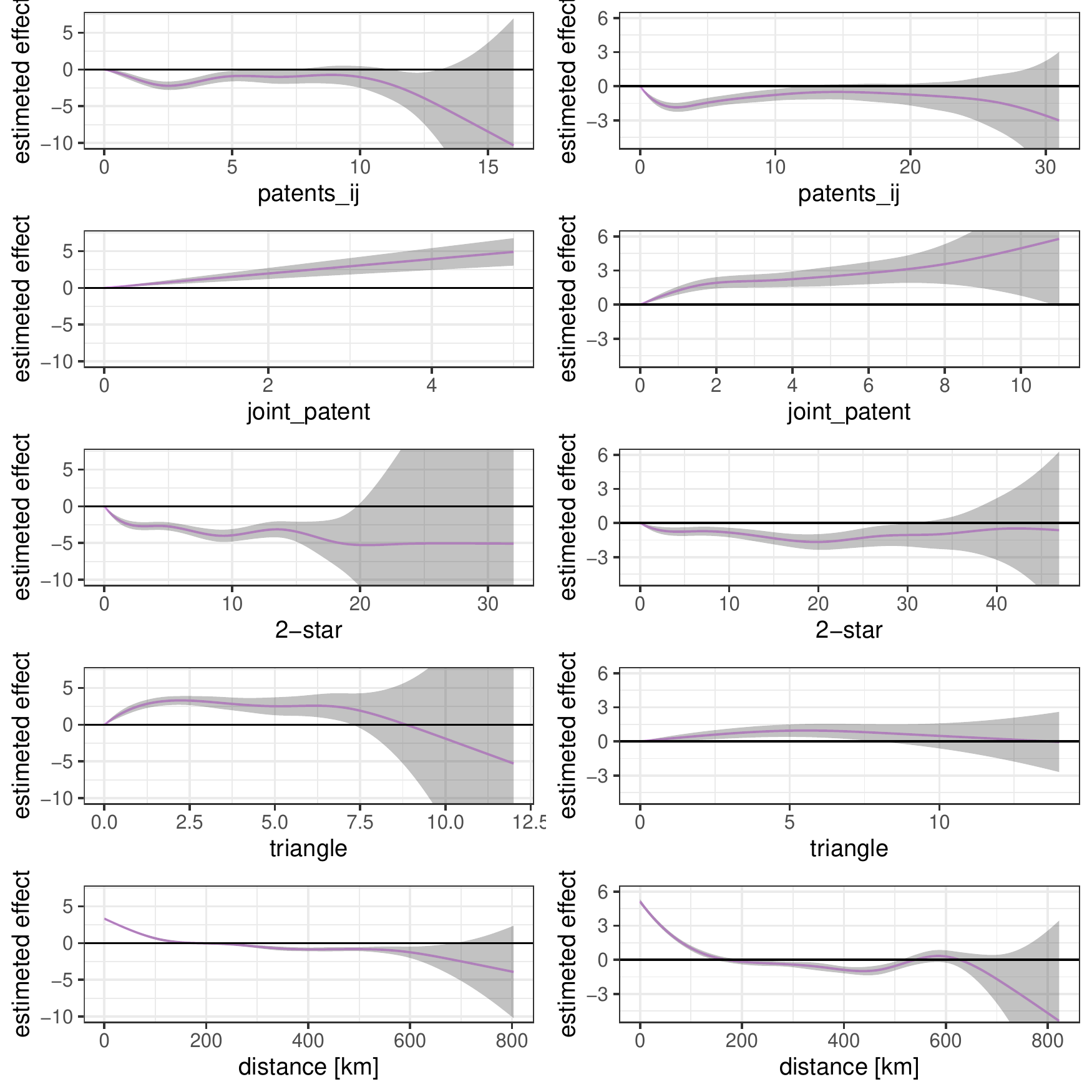}
\caption{Estimated smooth effects for \textit{IT-methods} (left panels) and \textit{food chemistry} (right panels) area and fourth time period.}
\label{Fig:all_smooth_effects_133_168_107_118}
\end{figure}

\clearpage
\section*{Appendix B: Technical Details}
\label{sec:technical_details}

The second-order difference penalty matrix can be defined as
\[\bm{K} = 
\begin{bmatrix}
\bm{K}_{(1)} & 0 & 0 & 0\\
0 & \bm{K}_{(2)} & 0 & 0 \\
0 & 0 & \dots & 0 \\
\vdots & \vdots & \vdots & \vdots \\
0 & 0 & 0 & \bm{K}_{(P)}\\
\end{bmatrix} \text{and}\;\;
\bm{K}_{(p)} = \begin{bmatrix} 
	   1 & -2 & 1 &  & & &   \\
   	  -2 & 5 & -4 & 1 & & &   \\
       1 & -4 & 6 & -4 & 1 & &  \\
         & \ddots & \ddots & \ddots & \ddots & \ddots &\\
      &  & 1 & -4 & 6 & -4 & 1   \\
      & & & 1 & -4 & 5 & -2\\
       & & & & 1 & -2 & 1 \\
\end{bmatrix}
\]
with dimension $\lbrack P\cdot K \times P\cdot K \rbrack$ and  $\lbrack K \times K \rbrack$, respectively. $P$ is the number of covariates. The second-order penalty matrix $\bm{K}$ can be derived from $\bm{K}_{(p)} = \bm{D}^T_2 \bm{D}_2$ where $\bm{D}_2 = \bm{D}_1 \bm{D}_{2-1}$ is a recursively obtained difference matrix with
\[\bm{D}_1 = 
\begin{bmatrix}
-1 & 1 &  & &  \\
 & -1 & 1 &  & \\
&  & \ddots &  \ddots & \\
& & & -1 & 1\\
\end{bmatrix}
\]
with dimension $\lbrack (K-1) \times K \rbrack$.
The corresponding derivatives to apply the Newton-Raphson algorithm are straight forward:
\begin{equation*}
s^{pen}(\bm{u}) = s(\bm{u}) - \left( \bm{K}(\bm{\lambda}) \right) \bm{u}
\end{equation*}
\begin{equation*}
J^{pen}(\bm{u}) = J(\bm{u}) - \bm{K}(\bm{\lambda})
\end{equation*}
%with $\bm{K}(\bm{\lambda}) = diag \left( \lambda_{(p)} \bm{K}_{(p)}\right)$ for $p = 1,\ldots P$, where $P$ is the number of covariates.

\clearpage

\end{appendix}
\clearpage
\noindent
{\bf{Acknowledgement} }

\noindent 
The project was partially supported by the European Cooperation in Science and Technology [COST Action CA15109 (COSTNET)].
\end{document}